\documentclass[12pt]{iopart}

\usepackage{iopams}
\usepackage{graphicx}
\usepackage{float}
\begin{document}

\title[]{Locally induced quantum interference in scanning gate experiments}

\author{A.~A.~Kozikov, R. Steinacher, C.~R\"{o}ssler, T.~Ihn, K.~Ensslin, C.~Reichl and W.~Wegscheider}

\address{Solid State Physics Laboratory, ETH Z\"{u}rich, CH-8093 Z\"{u}rich, Switzerland}
\ead{akozikov@phys.ethz.ch}
\begin{abstract}

We present conductance measurements of a ballistic circular stadium influenced by a scanning gate. When the tip depletes the electron gas below, we observe very pronounced and regular fringes covering the entire stadium. The fringes correspond to transmitted modes in constrictions formed between the tip-induced potential and the boundaries of the stadium. Moving the tip and counting the fringes gives us exquisite control over the transmission of these constrictions. We use this control to form a quantum ring with a specific number of modes in each arm showing the Aharonov-Bohm effect in low-field magnetoconductance measurements.

\end{abstract}


\maketitle

\section{Introduction}

Semiconductor nanostructures are usually defined electrostatically using top-down or bottom-up approaches. For example, one can use suitably biased lateral gates or self-assembled systems based on sophisticated growth schemes to prepare semiconductor quantum structures. In all these cases the main features of the potential landscape are defined for each sample and only weakly tunable. Using the conductive tip of a scanning force microscope as a movable top gate allows for more advanced control over the potential. In addition to the tip-sample bias voltage, the tip-surface distance and the in-plane tip position can be changed. Variable strength and gradient of the tip-induced potential give the possibility to tailor the potential landscape of nanostructures. In the past this scanning gate microscopy (SGM) technique was used to study the conductance of predefined nanostructures. For example, Topinka et al., Paradiso et al., Kozikov et al. imaged electron backscattering through a quantum point contact \cite{TopinkaSci, TopinkaNat, Paradiso, KozikovBranches}. These experiments demonstrated that SGM allows observing features which are smaller than the Fermi wavelength. In addition,  Pioda et al., Fallahi et al. looked into single electron transport in quantum dots \cite{Pioda2004,Fallahi} and Woodside et al. in carbon nanotubes \cite{Woodside}. Hackens et al. probed electron transport through a quantum ring \cite{Hackens,Hackens1}.

In this work we demonstrate how to extend this technique to alter the potential landscape in a ballistic stadium using a scanning gate. Previously the conductance was studied in ballistic stadii with chaotic dynamics of electrons \cite{Crook,Aoki}. In contrast to their results, we observe regular modulations and fringe patterns in the spatially resolved conductance covering the entire structure. The achieved control over the potential landscape is used to form a quantum ring showing Aharonov-Bohm oscillations. This requires to have a spatially and temporally very stable setup. The observation of the Aharonov-Bohm oscillations in our samples testifies to the quality of our experimental setup as well as to the potential to investigate quantum effects in more detail in complex nanostructures whose potential can be tailored with the scanning gate.

\section{Experimental methods}
The microstructure is fabricated on a high-mobility GaAs/AlGaAs heterostructure. The electron gas buried 120 nm beneath the surface has a mobility of 3.8$\times 10^{6}$ cm$^2/$Vs at 300 mK at a carrier density of 1.5$\times 10^{11}$ cm$^{-2}$. This gives an elastic mean free path $l_\mathrm{e}=$ 50 $\mu$m and a Fermi wavelength of electrons $\lambda_\mathrm{F}=$ 65 nm.

\begin{figure}[H]
\begin{center}
\includegraphics[width=15cm]{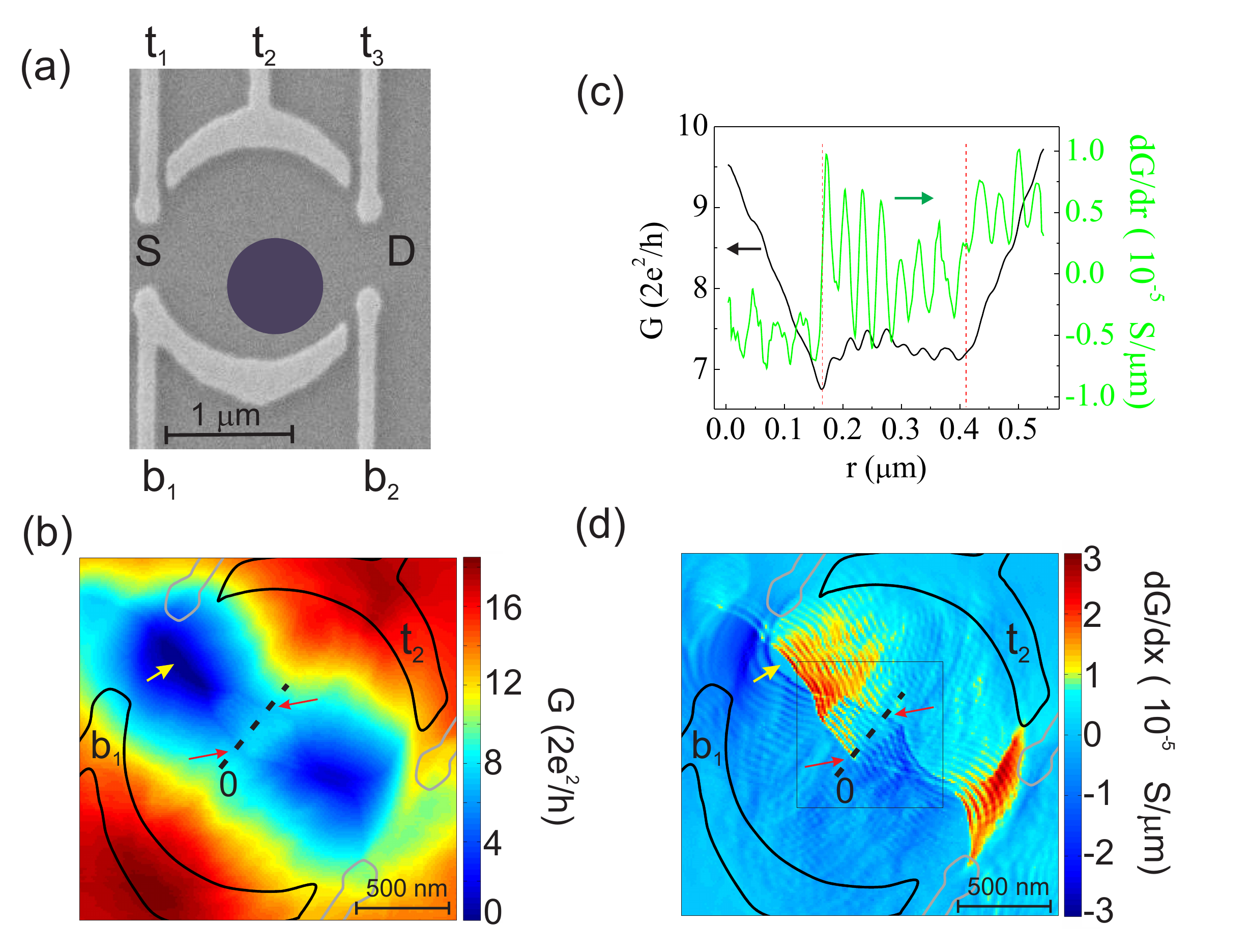}
\caption{(a) SEM image of the device. The dark grey area corresponds to the GaAs surface, bright grey regions labeled t$_1$, t$_2$ and t$_3$, b$_1$ and b$_2$ are metallic top gates. The ac current flows between the source ``S'' and drain ``D'' contacts. The size of the tip-depleted disc is schematically represented by the dark blue circle. (b) Conductance, $G$, through the stadium in units of 2e$^2$/h as a function of tip position, $(x,y)$. The biased t$_2$ and b$_1$ and grounded t$_1$, t$_3$ and b$_2$ top gates are outlined by black and grey solid lines, respectively. (c) Conductance and its numerical derivative along the black dashed line shown in (b). Vertical red dashed lines in (c) together with red arrows in (b) mark the borders of the checker-board pattern. The coordinate $r=0~\mu$m is labeled at the left end of the black dashed line. (d) Numerical derivative of the conductance in (b), $\rmd G(x, y)/\rmd x$, as a function of tip position. The black dashed line is at the same position as that in (b). The black square is the region where high-resolution measurements are taken. Yellow arrows in (b) and (d) mark the lens-shaped region of zero conductance arising when the tip is close to the left constriction. Voltages -0.5 and -4.0 V are applied to the top gates and the tip, respectively. The tip is placed 60 nm above the GaAs surface. Biased top gates are labeled in (b) and (d).}
\label{fig:LargeStadium}
\end{center}
\end{figure}

The device under study consists of three circular stadii connected in series with different lithographic diameters 1.0, 1.2 and 1.5 $\mu$m. In this paper we focus on the largest stadium with the diameter $D$ = 1.5 $\mu$m, which is shown in \fref{fig:LargeStadium}(a). The structure is defined by applying a negative voltage to the corresponding metallic gate electrodes t$_1$, t$_2$, t$_3$,  b$_1$ and b$_2$ fabricated by e-beam lithography, each 30 nm high. The electron gas beneath them becomes depleted at -0.4 V. Since $D\ll l_\mathrm{e}$, transport through the stadium is ballistic. At the same time $D\gg \lambda_\mathrm{F}$.

The measurements are performed in a $^3$He system with a base temperature of 300 mK using a home-built scanning force microscope \cite{Ihn}.
The two-terminal linear conductance $G$ through the device is measured by applying a 26 Hz ac rms voltage of 100 $\mu$V between the source and drain contacts. Scanning gate measurements are performed by placing the metallic biased tip of the scanning force microscope 60 nm above the sample surface. The tip scans the surface at a constant height and $G$ is recorded simultaneously leading to 2D maps $G(x,y)$ as a function of lateral tip position $(x,y)$. A voltage of -4 V between the tip and the 2DEG is chosen to deplete the electron gas beneath the tip. The resulting diameter of the depletion region is about 0.7 $\mu$m \cite{KozikovStadium}.

\Fref{fig:LargeStadium}(b) shows the conductance through the stadium in color when a voltage of -0.5 V is applied to the top gates t$_2$ and b$_1$ outlined with black solid lines. This voltage reduces the electronic size of the structure by about 100 nm compared to the geometric size.
When the tip is inside the structure, $G$ decreases from the tip-unpertubed value $G\approx17.2\times2e^2/h$ to zero (lens-shaped regions of dark blue color marked by the yellow arrow). Such a drop of the conductance is due to backscattering of electrons at the tip-depleted region. As the tip comes close enough to one of the constrictions, its depletion disc blocks the current entirely. The conductance is not reduced to zero when the tip is close to the right constriction, because this constriction is larger than the left. When the tip is above the top gates or outside the structure, the conductance is enhanced to $G\approx 18.5\times2e^2/h$ (dark red area) compared to $G$ without the tip. Such an enhancement has been observed by us in seven different stadii, four of which have different diameters. Qualitatively it is due to a tip-induced transition from Ohmic to adiabatic transport \cite{KozikovStadium, QPCseries}.

On closer inspection one can see faint fringes in the central region (along the dashed line between two red arrows) of the stadium. \Fref{fig:LargeStadium}(c) shows the corresponding conductance along the black dashed line in (b). Indeed, $G$ oscillates in this region (enclosed between the two vertical red lines) and stays on average roughly at a constant value of $G\approx7\times2e^2/h$. Such behavior can intuitively arise from the formation of two narrow channels between the tip-depleted region and t$_2$ and b$_1$ each having its own quantized conductance $G_1$ and $G_2$. Classically the total conductance would then be $G=G_1+G_2$. Outside the central region one of these channels is depleted and $G$ is determined by the remaining channel, i.e. either $G_1$ or $G_2$. As this channel becomes wider, $G$ increases steeply as seen in the two regions outside the vertical dashed red lines. Small ``shoulders" in $G$ seen in these two regions correspond to quantized conductance of the remaining channel. Such small modulation of $G$ can be revealed by taking a numerical derivative $\rmd G(r)/\rmd r$ (green curve in \fref{fig:LargeStadium}(c)). Conductance oscillations are now clearly seen along the entire length of the chosen 1D cut.

To reveal such small changes in $G$ in the entire area of \fref{fig:LargeStadium}(b), a numerical derivative $\rmd G(x, y)/\rmd x$ with respect to the scan direction is plotted in \fref{fig:LargeStadium}(d). Several fringe patterns covering the entire structure are now visible. Two of them are located at the left constriction around the lens-shaped region (marked by the yellow arrow) and at the right constriction. These fringes originate from conductance quantization in single QPCs formed between the tip-induced potential and the boundaries of the stadium \cite{KozikovStadium}. For example, when the tip moves from the left lens-shaped region towards the top gate t$_2$, a constriction opens gradually between the tip and the lower top gate b$_1$. As its width increases, more modes are transmitted through it. Each added mode corresponds to one fringe period.

Each fringe at the left lens-shaped region in \fref{fig:LargeStadium}(d) evolves continuously through the center of the stadium to a corresponding fringe at the right constriction. This gives rise to a checkerboard pattern formed in the central region of the structure. The 1D cut discussed in (c) is taken across this pattern (its borders are marked by the same red arrows as in (b)).

\begin{figure}[H]
\begin{center}
\includegraphics[width=10cm]{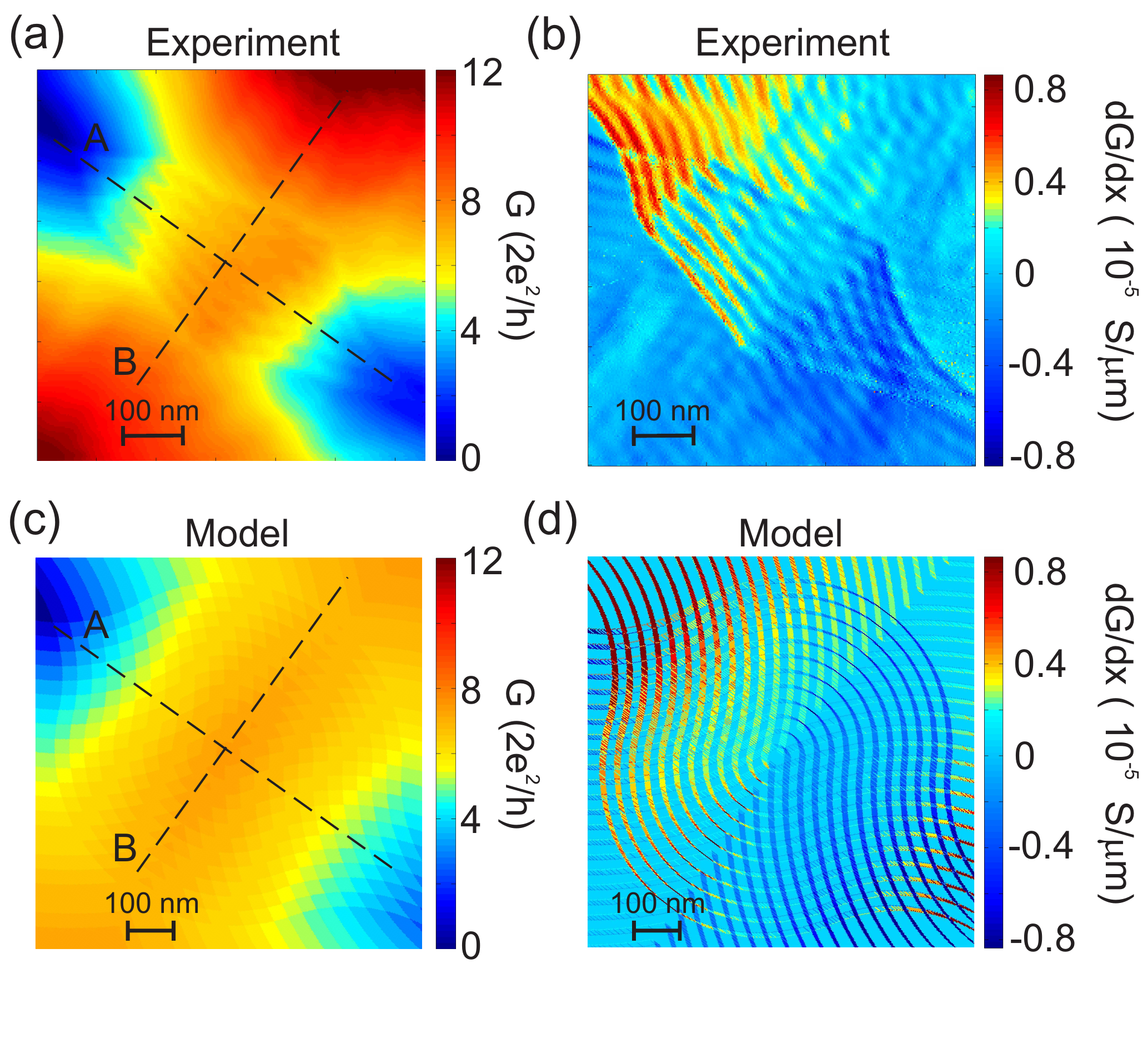}
\caption{(a) Conductance, $G(x,y)$, through the stadium 1.5 $\mu$m in diameter in units of 2e$^2$/h as a function of tip position in the region indicated in \fref{fig:LargeStadium}(d) by a square. (b) Numerical derivative of the conductance in (a), $\rmd G(x, y)/\rmd x$. The top gates and the tip are biased to -0.5 and -4.0 V, respectively. (c) Modeled conductance and (d) its numerical derivative.}
\label{fig:LargeStadiumZoom}
\end{center}
\end{figure}

To check the proposed origin of the checkerboard pattern in the center of the stadium, we take a separate high-resolution measurement of this region and compare the results with computer modeling. \Fref{fig:LargeStadiumZoom} shows the mentioned checkerboard pattern both in the raw conductance (a) and in its numerical derivative (b). In order to model this situation, we consider four constrictions: two, a and b, at the entrance and exit of the stadium. The two others, c and d, form between the tip-depleted region and either gate t$_2$ or $b_1$ \cite{KozikovStadium}. We treat them as four incoherently coupled conductors. The total conductance is then $G^{-1}=G_\mathrm{a}^{-1}+\left( G_\mathrm{c}+G_\mathrm{d} \right)^{-1}+G_\mathrm{b}^{-1}$. The conductance of each constriction is taken to be $G_i=\left(2e^2/h\right)\times N_i$ ($i$ = a, b, c and d), where $N_i=2W_i/\lambda_\mathrm{F}$ is the number of transmitted modes in constriction $i$ and $W_i$ is the width of constriction $i$. As the tip moves, only $N_\mathrm{c}$ and $N_\mathrm{d}$ change in our model for simplicity. The numbers $N_\mathrm{a}=18$ ($W_\mathrm{a}=0.6~\mu$m) and $N_\mathrm{b}=25$ ($W_\mathrm{b}=0.8~\mu$m) are kept fixed. The resulting $G$ and $\rmd G(x, y)/\rmd x$ are shown in \fref{fig:LargeStadiumZoom}(c) and (d), respectively, as a function of tip position.

\begin{figure}[H]
\begin{center}
\includegraphics[width=15cm]{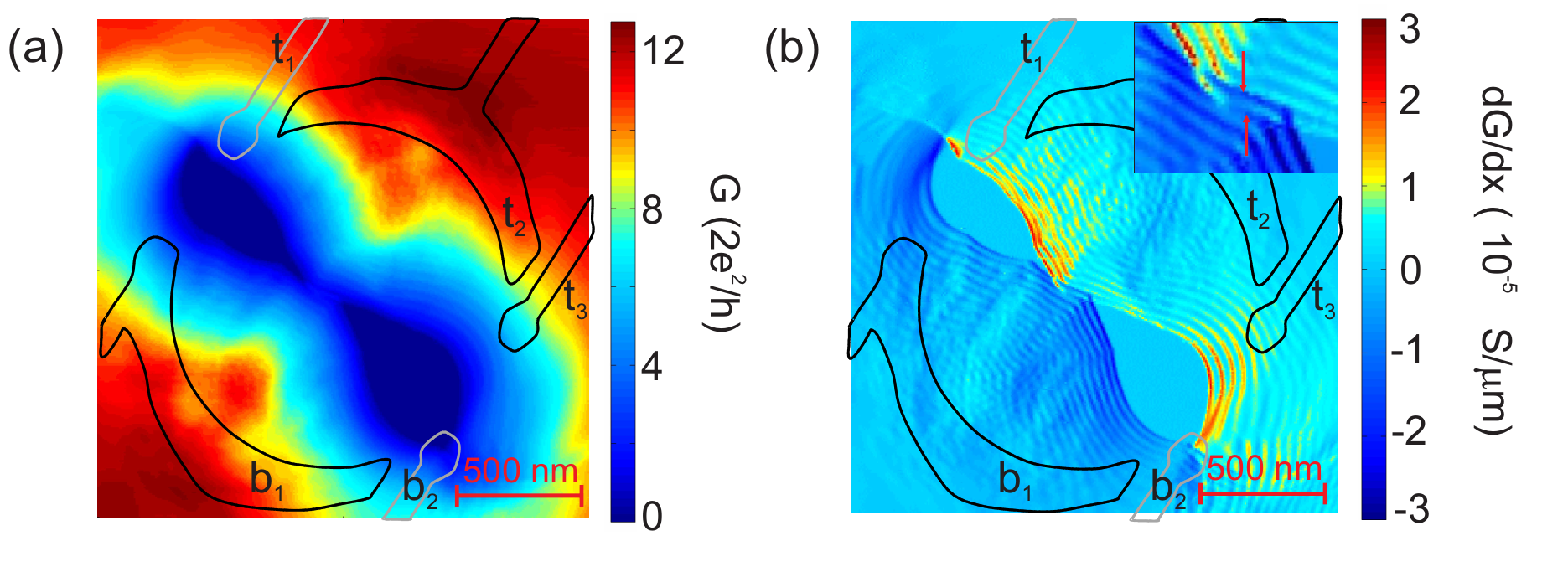}
\caption{(a) Conductance, $G(x,y)$, through the largest stadium. The top gates are now biased (black solid lines) to form a more symmetric stadium. The grounded top gates are indicated by grey solid lines. (b) Numerical derivative of the conductance in (a), $\rmd G(x, y)/\rmd x$. The inset is the central region of the stadium enlarged for clarity. The red arrows indicate two fringes each corresponding to one transmitted mode in each arm of the ring. All top gates are labeled.}
\label{fig:LargeStadiumAB}
\end{center}
\end{figure}

The model shows a reasonable qualitative and quantitative agreement with the experimental data along line A shown in \fref{fig:LargeStadiumZoom}(a) and (c). Along line B the agreement is only qualitative. This is because the model does not take into account the tip-induced transition from Ohmic to adiabatic transport. In addition, the tip-induced potential in the model is assumed to be hard-wall and the four conductors are incoherently coupled. The latter is a rough approximation.
When the tip is in the central part of the stadium, it forms an electronic ring together with the boundaries of the cavity (gates t$_2$ and b$_1$). Since each arm gives its own set of parallel fringes, a checkerboard pattern forms in the model in \fref{fig:LargeStadiumZoom}(d) similar to the experimental observation in \fref{fig:LargeStadiumZoom}(b). The shape of the fringes follows the inner edge of the respective top gate. The experiment shows rather sharp kinks in the fringe pattern, whereas the simulation does not. This could arise, because the tip-induced potential or the electrostatic edges of the stadium may not be perfectly round.

Counting the measured fringes enables us to determine and then set the exact number of transmitted modes in the constrictions formed between the tip-induced potential and the boundaries of the stadium. Having such an exquisite control allows us to tune the system to form an Aharonov-Bohm ring with a specific number of modes in each arm.
As an example, in \fref{fig:LargeStadiumAB} voltages are applied to the gates b$_1$, t$_2$ and t$_3$ to have a symmetric stadium (notice the almost same size of the lens-shaped regions) with only two fringes in its middle (see inset of \fref{fig:LargeStadiumAB}(b)). The number of fringes in the checkerboard pattern yields the number of transmitted modes in each arm of the ring. Placing the tip between these two fringes (inset in \fref{fig:LargeStadiumAB}(b)) and in the middle between the lens-shaped regions forms the desired Aharonov-Bohm (AB) ring. We tune the entrance and exit of this ring to transmit only a few modes. For this purpose, after fixing the tip position as described, the two previously grounded gates t$_1$, b$_2$ are biased to the same voltage of -0.5 V. The conductance, $G(B, V_\mathrm{g})$, is then measured as a function of the magnetic field and the voltage applied to all top gates of the structure.

\begin{figure}[t]
\begin{center}
\includegraphics[width=15cm]{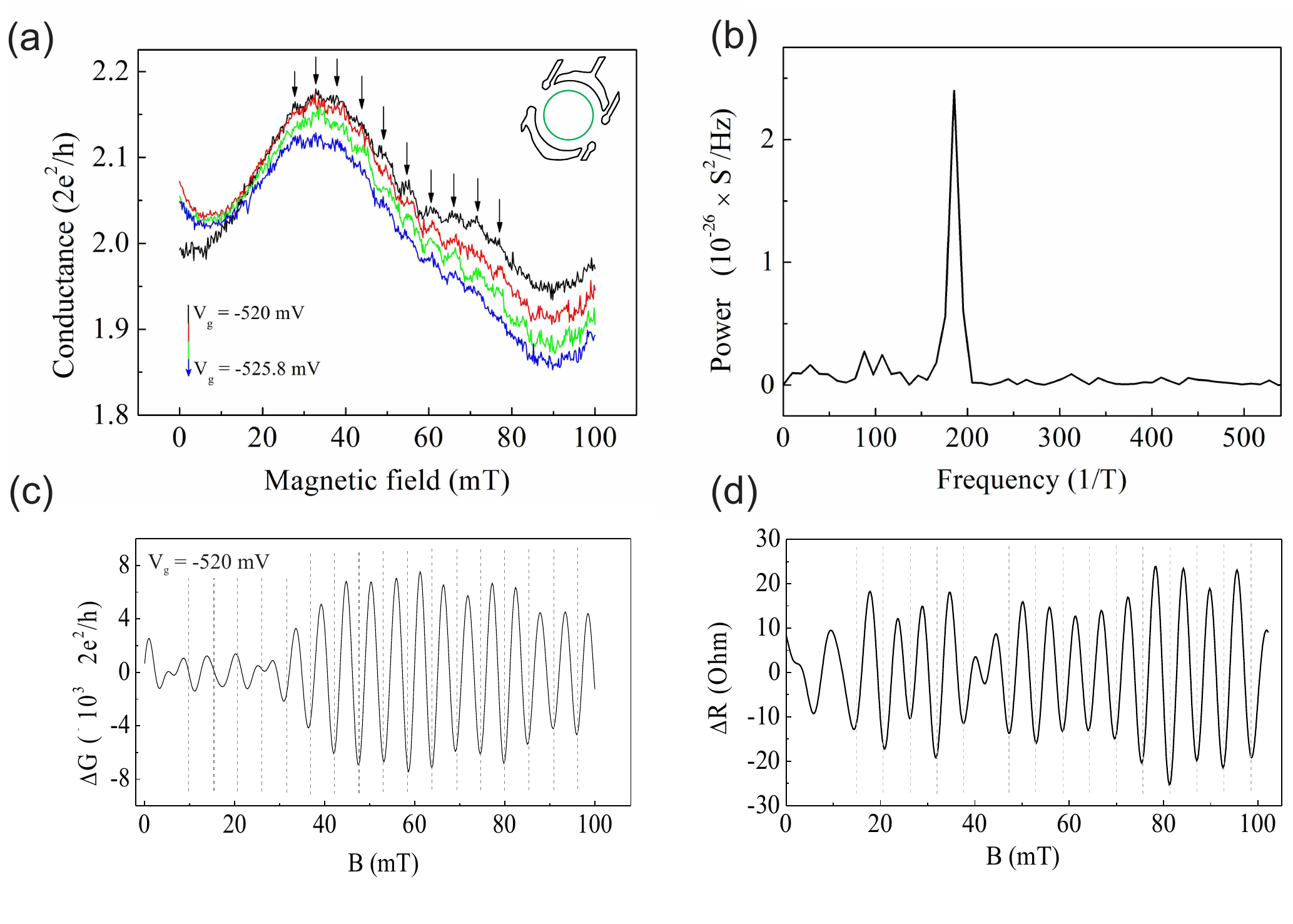}
\caption{(a) Low-field magnetoconductance measured through a tip formed ring. Curves of different colors correspond to different gate voltages with a step of -1.93 mV. The inset shows a schematics of the tip-formed Aharonov-Bohm ring. (b) A Fourier spectrum of the black curve in (a) with the subtracted background. (c) and (d) Filtered data from 2- and 4-terminal measurements, respectively. Equidistantly spaced vertical dashed lines are guides to the eye.}
\label{fig:ABoscillations}
\end{center}
\end{figure}

The resulting $G(B,V_\mathrm{g})$ taken at a source-drain bias of 10 $\mu$V are presented in \fref{fig:ABoscillations}(a). They show a small (about 0.3$\%$) modulation of $G$ marked by black arrows. Subtracting a slowly varying background due to classical effects and taking a Fourier spectrum reveals a single peak at a frequency of about 185 T$^{-1}$ (\fref{fig:ABoscillations}(b)). This peak corresponds to an oscillation period of 5.4 mT, which agrees roughly with the $h/e$ period of a ring with an estimated diameter of 1 $\mu$m. One conductance curve band-pass filtered around the peak in the Fourier spectrum is shown in \fref{fig:ABoscillations}(c).

The AB oscillations are found to be very sensitive to the tip position on the scale of half the Fermi wavelength. They exist only in a narrow range of gate voltages. Nevertheless, the oscillations were reproducibly found in different cooldowns and with different tips measured in 2- and 4-terminal configuration.
As an example, we present in \fref{fig:ABoscillations}(d) the result of the filtered data similar to those shown in \fref{fig:ABoscillations}(c) but obtained using a four-terminal configuration by passing a current of 1 nA through the sample. The relative amplitudes of the AB oscillations in two- (about 0.3$\%$) and four-terminal (about 0.5$\%$) measurements are very similar.


\section{Conclusion}

We have presented results of transport measurements through a ballistic circular stadium, the potential of which was modified by a scanning gate. Conductance maps as a function of tip position revealed fringe patterns covering the entire structure. The fringes correspond to conductance quantization plateaus in constrictions formed between the tip-induced potential and the boundaries of the stadium. At the entrance and exit of the cavity the tip formed single constrictions together with either gate of the stadium. This lead to single sets of fringes. In the center of the microstructure they crossed each other forming a checkerboard pattern. That corresponded to a situation when the tip created a quantum ring. Transport in this case occurred through both arms of the ring. Results of computer modeling agreed qualitatively with our experimental observations.

To transmit through the stadium, electron waves split into two by the tip and interfere when recombined at the other side. By counting the fringes a specific number of transmitted modes in the arms of the ring was set. As a result we observed conductance oscillations as a function of low magnetic field due to the Aharonov-Bohm effect.

\section{Acknowledgements}

We are grateful for fruitful discussions with Fabrizio Nichele. We acknowledge financial support from the Swiss National Science Foundation and NCCR ``Quantum Science and Technology".


\newpage

\section*{References}

\begin{thebibliography}{99}
\bibitem{TopinkaSci} Topinka M A, LeRoy B J, Shaw S E J, Heller E J, Westervelt R M, Maranowski K D and Gossard A C 2000 {\it Science} {\bf 289} 2323
\bibitem{TopinkaNat} Topinka M A, LeRoy B J, Westervelt R M, Shaw S E J, Fleischmann R, Heller E J, Maranowski K D and Gossard A C 2001 {\it Nature} {\bf 410} 183
\bibitem{Paradiso} Paradiso N, Heun S, Roddaro S, Pfeifer L N, West K W, Sorba L, Biasiol G, Beltram F 2010 {\it Phisica E} {\bf 42} 1038
\bibitem{KozikovBranches} Kozikov A A, R\"{o}ssler C, Ihn T, Ensslin K, Reichl C and Wegscheider W 2013 {\it New J. Phys.} {\bf 15} 013056
\bibitem{Pioda2004} Pioda A, Kicin S, Ihn T, Sigrist M, Fuhrer A, Ensslin K, Weichselbaum A, Ulloa S E, Reinwald M and Wegscheider W 2004 {\it Phys. Rev. Lett.} {\bf 93} 216801
\bibitem{Fallahi} Fallahi P, Bleszynski A C, Westervelt R M, Huang J, Walls J D, Heller E J, Hanson M and Gossard A C 2005 {\it Nano Lett} {\bf 5} 223
\bibitem{Woodside} Woodside M T, McEuen P L 2002 {\it Science} {\bf 296} 1098
\bibitem{Hackens} Hackens B \textit{et al.} 2006 {\it Nature Physics} {\bf 2} 826
\bibitem{Hackens1} Hackens B \textit{et al.} 2010 {\it Nat. Commun.} {\bf 1} 39
\bibitem{Crook} Crook R, Smith C G, Graham A C, Farrer I, Beere H E, and Ritchie D A 2003 {\it Phys. Rev. Lett} {\bf 91} 246803
\bibitem{Aoki} Aoki N, Brunner R, Burke A M, Akis R, Meisels R, Ferry D K, and Ochiai Y 2012 {\it Phys. Rev. Lett} {\bf 108} 136804
\bibitem{Ihn} Ihn T, ``Electronic Quantum Transport in Mesoscopic Semiconductor Structures", Springer Tracts in Mod. Phys. 192, (2004)
\bibitem{KozikovStadium} Kozikov A A, D. Weinmann, R\"{o}ssler C, Ihn T, Ensslin K, Reichl C and Wegscheider W 2013 {\it New J. Phys.} {\bf 15} 083005
\bibitem{QPCseries} Kouwenhoven L P, van Wees B J, Kool W, Harmans C J P M, Staring A A M, Foxon C T 1989 {\it Phys. Rev. B} {\bf 40} 8083


\end{thebibliography}
\end{document}